# Comparative Evaluation of Transition Mechanisms for Adaptive Droop Gains in Parallel Grid-Forming Inverters

Edgar Diego Gomez Anccas*, Eilean Angus MacPherson, Jirko Tegeler,
Detlef Schulz
Chair of Electrical Power Systems, Helmut Schmidt University, Hamburg, Germany
*Corresponding author: diego.gomez@hsu-hh.de

***Abstract*—Uncertainty in standalone microgrid operation usually originates from mismatches between power references and forecasts. These deviations are compensated by grid-forming controlled units, which distribute the required power contribution based on their droop gains. To introduce an additional degree of flexibility, it is possible to treat droop gains as decision variables to redistribute active-power contributions according to system-level objectives. However, directly applying updated droop gain references from a supervisory layer to the primary controllers can introduce power and frequency transients. This paper investigates transition mechanisms for applying scheduled active-power droop gain changes during operation. Hard switching, rate-limited transition, first-order IIR low-pass filtering, and cubic as well as quintic S-curve transitions are compared experimentally on two parallel 15 kW grid-forming inverter units. The results show that shaping the droop gain trajectory significantly reduces transient deviations compared to hard switching. In the considered case study, the S-curve transitions provide the strongest transient mitigation, reducing the active-power overshoot from $632.7\,\mathrm{W}$ to approximately $115\,\mathrm{W}$ and limiting the frequency overshoot to about $0.003\,\mathrm{Hz}$.**

## I. Introduction

In modern power systems, new challenges, including infeed from renewable energy sources (RES) such as photovoltaic (PV) generators and wind turbines, as well as changes in demand profiles due to penetration of electric vehicles arise [1], [2]. Moreover, this development entails a shift from centralised operation based on large conventional generators towards a more distributed, power electronics-based energy system. For standalone systems such as islanded microgrid (MG) in particular, where access to large reserves, rotary inertia or power exchange is limited, advanced digitalisation and control are required [3].

To operate these systems reliably and economically, a holistic evaluation of all control layers operating across different time domains is imperative. Two key layers are the operation control and the primary control layer. At the operational control layer, optimal power references for the lower control layers are calculated while ensuring safe operation bounds and maximising renewable utilisation. For these tasks, model predictive control (MPC) is a promising approach, where an optimisation problem is solved following a receding horizon method while accounting for uncertainty and system constraints.

In the primary control layer of inverter-dominated microgrids, a distinction is made between grid-following and grid-forming inverter (GFMI) due to their behaviour as current and voltage sources, respectively [4]. In standalone microgrids, grid-forming control is usually employed to establish voltage and frequency and to compensate power imbalances. While the implementation is straightforward for a single unit, once multiple GFMI are operated in parallel active and reactive power sharing needs to be addressed. Depending on the controller implementation the sharing requires additional communication [5], [6]. The link between both layers is realised through setpoint distribution and consideration of grid-forming power sharing in the optimisation problem [7]. Furthermore, adaptive droop gains allow the shaping of fluctuation handling and can therefore increase renewable utilisation in standalone microgrids [8]. Several studies have addressed adaptable droop gains for example to enhance damping and stability [9], to balance energy storage state-of-charge [10], to improve efficiency in operation [11] or to reduce power losses during transmission [12]. Only a few contributions explicitly address the transition of droop gains. In [13], gain changes are smoothed using rate limiters and dead-bands.

Motivated by these contributions, this work provides an experimental assessment of different transition methods for adaptable droop gains. The gains are derived from a robust MPC scheme and applied to the primary controllers of parallel GFMI in a low-voltage, laboratory-scale microgrid. Their impact is assessed in terms of transient behaviour, power-sharing performance, and stable operation. This work is structured as follows. In Section II the link between the two control layers is introduced. Section III presents the adaptive power sharing and transition mechanisms. Section IV introduces the experimental setup and presents the comparison results. Lastly, Section V provides an outlook on future work.

This research paper is funded by dtec.bw–Digitalization and Technology Research Center of the Bundeswehr–which we gratefully acknowledge. dtec.bw is funded by the European Union–NextGenerationEU.

## II. MPC-Scheduled Adaptive Power Sharing

A complete MPC problem formulation for microgrids generally includes forecasts for load and renewable infeed, operational constraints, an objective function, a description of uncertainty and a microgrid model. For application to a real system, the computed optimal inputs affect control layers with different time scales. In this work, the two considered layers satisfy

$$T_{\mathrm{mpc}} > T_{\mathrm{pc}}, \tag{1}$$

where $T_{\mathrm{mpc}}$ is the sample time of the supervisory MPC and $T_{\mathrm{pc}}$ the sample time of the primary controller. In this case, a minmax MPC formulation is considered, where uncertain input $w(k)$ is assumed to be bounded by

$$w(k) \in [\underline{w}(k), \overline{w}(k)]\,. \tag{2}$$

The uncertain quantities considered by the supervisory layer affect the power balance of the microgrid and therefore the power contributions required from the grid-forming units. This sharing behaviour is governed by the active-power droop characteristics of the grid-forming controllers. If the MG exhibits synchronized motion [14], then

$$0 = \dot{\theta}_i(t) - \dot{\theta}_j(t), \tag{3a}$$

$$\omega_i(t) = \omega_j(t) = \omega(t), \tag{3b}$$

$$\dot{p}_i^{\mathrm{m}}(t) = 0. \tag{3c}$$

Consequently,

$$p_i^{\mathrm{m}}(t) = p_i(t), \tag{4}$$

Here, $\theta$ is the phase angle, $\omega$ the frequency and $p^m$ the power measurements. Proportional power sharing among grid-forming units described in discrete time is then achieved according to

$$\left(p_i(k) - u_i(k)\right) K_{pi} = \left(p_j(k) - u_j(k)\right) K_{pj}. \tag{5}$$

Hence, the droop-gain ratio between two units determines the ratio of the active-power deviation covered by each unit. Instead of keeping the droop gains fixed, they can be treated as decision variables $K_{pi}(k)$ and $K_{pj}(k)$ in an MPC formulation. This introduces an additional degree of freedom for coordination, since the contribution of each unit can be adapted according to the optimisation objective and the operational constraints. This results in the computed optimal droop reference $K_{pi}^*(k)$, which is transmitted to the primary control at a sample time of $T_{\text{mpc}}$. However, directly applying a new value of $K_{pi}^*$ to the primary droop controller may introduce discontinuities in the frequency reference and excite transient power oscillations. Therefore, the reference droop gain must be transferred to the applied droop gain in a controlled manner.

In Figure 1 the complete segment is shown with the MPC generating $K_{pi}^*(k)$ at a sample time $T_{\text{mpc}}$ based on the forecast bounds $\left[\underline{\hat{w}}(k), \overline{\hat{w}}(k)\right]$. Next, the reference is conditioned before being transmitted to the primary controller, to smoothen the system behaviour. The primary control layer generates voltage and angle references $v^*(k)$, $\delta^*(k)$, respectively for the microgrid, which in turn feeds back voltage, current and power measurement values $v(k)$, $i(k)$, $p(k)$ for the primary controler and state $x(k)$ and input data $v(k-1)$ for the MPC.

The following section investigates different transition mechanisms for this mapping. The objective is to apply scheduled droop-gain changes during operation while limiting frequency deviations and active-power transients.

## III. Transition Mechanisms for Adaptive Power Sharing

Adapting the droop gain of an operating GFMI changes its droop characteristics and may result in transients, even if the system is initially in steady state. The simplest implementation is to directly apply the optimal gain reference to the primary controller. Hence, at time step $k$

$$K_{p1}(k) = K_{p1}^*(k), \tag{6}$$

holds. Figure 2 shows the effects of instantaneous switching compared to a rate-limited transition. The switching case exhibits larger peak deviations and more pronounced oscillations. This motivates the investigation of smoother transition mechanisms for GFMIs.

### A. Rate-Limited Transition

In the rate-limited transition, the applied droop gain is moved towards the reference value $K_{p1}^*(k)$, while the maximum step size is limited to $mT_{\text{pc}}$ at every primary controller

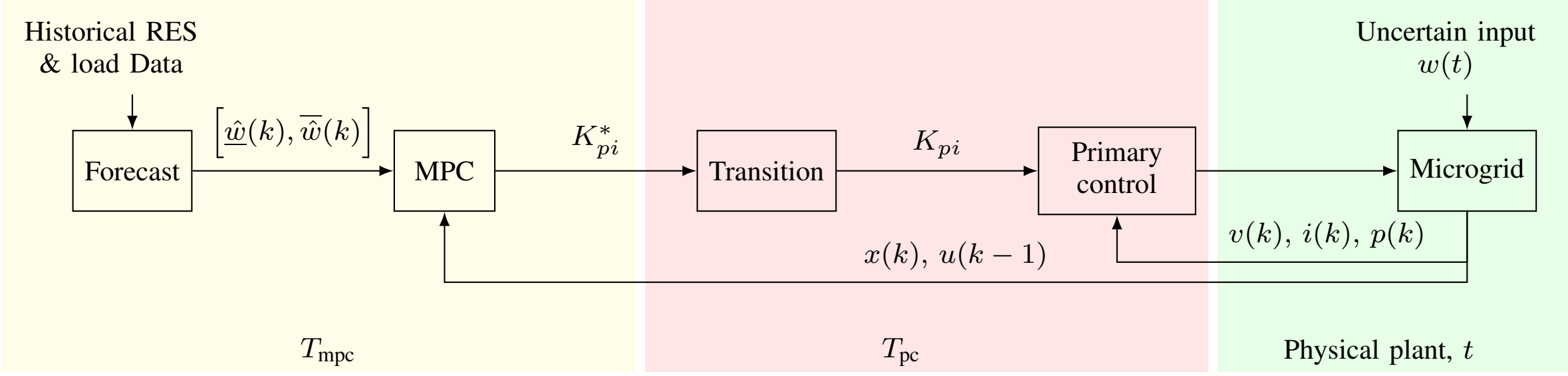

Fig. 1. Supervisory scheduling of the reference droop gain by the MPC layer.

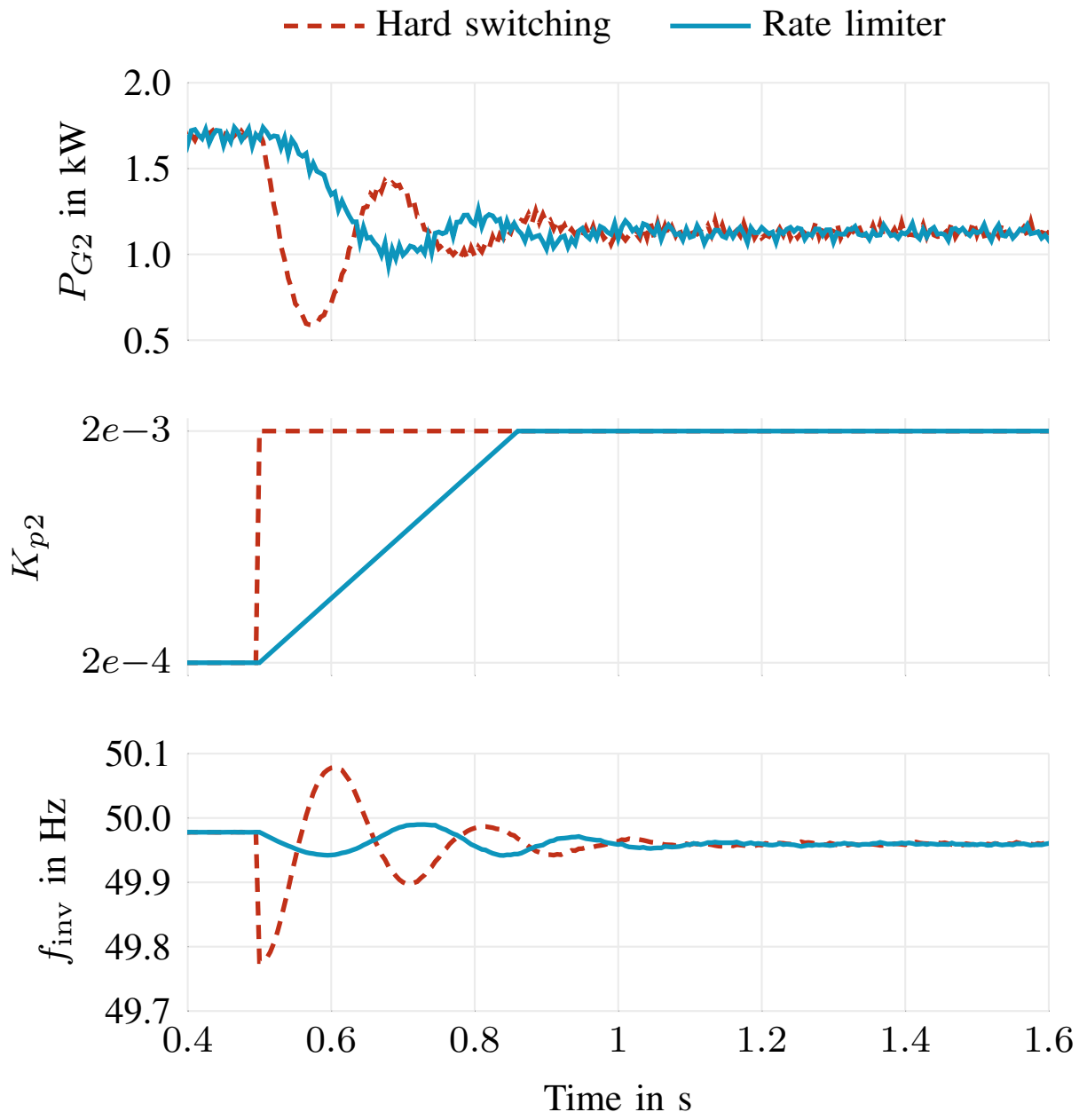

Fig. 2. Comparison of instantaneous switching and rate-limited transition of the droop gain during parallel operation of two grid-forming units.

time step. With $e_K(k) = K^*_{p1}(k) - K_{p1}(k)$, the update law is given by

$$K_{p1}(k+1) = \begin{cases} K_{p1}(k) + mT_{\text{pc}}, & e_K(k) > mT_{\text{pc}}, \\ K_{p1}(k) - mT_{\text{pc}}, & e_K(k) < -mT_{\text{pc}}, \\ K^*_{p1}(k), & |e_K(k)| \leq mT_{\text{pc}}. \end{cases} \tag{7}$$

Here, $m$ denotes the maximum admissible rate of change of the droop gain.

### B. First-Order IIR Low-Pass Filter Transition

An alternative method for smoothing the transition is to filter the reference droop gain using a first-order IIR low-pass filter. The applied droop gain at time step k is updated according to

$$K_{p1}(k+1) = K_{p1}(k) + \alpha(k) e_K(k). \tag{8}$$

For a cutoff frequency $f_c$, the filter coefficient is computed as

$$\alpha(k) = \frac{2\pi f_c T_{\text{pc}}}{1 + 2\pi f_c T_{\text{pc}}}. \tag{9}$$

Thus, the transition dynamics can be adjusted through the cutoff frequency. A higher cutoff frequency accelerates the tracking of the reference value, and vice versa.

### C. S-Curve Transition

For the S-curve transition, changes in the reference droop gain $K^*_{p1}(k)$ are applied through a smooth interpolation. Once $K^*_{p1}(k) \neq K^*_{p1}(k-1)$, the current applied gain and the new reference value are stored as

$$K^0_{p1} = K_{p1}(k), \qquad K^{*0}_{p1} = K^*_{p1}(k). \tag{10}$$

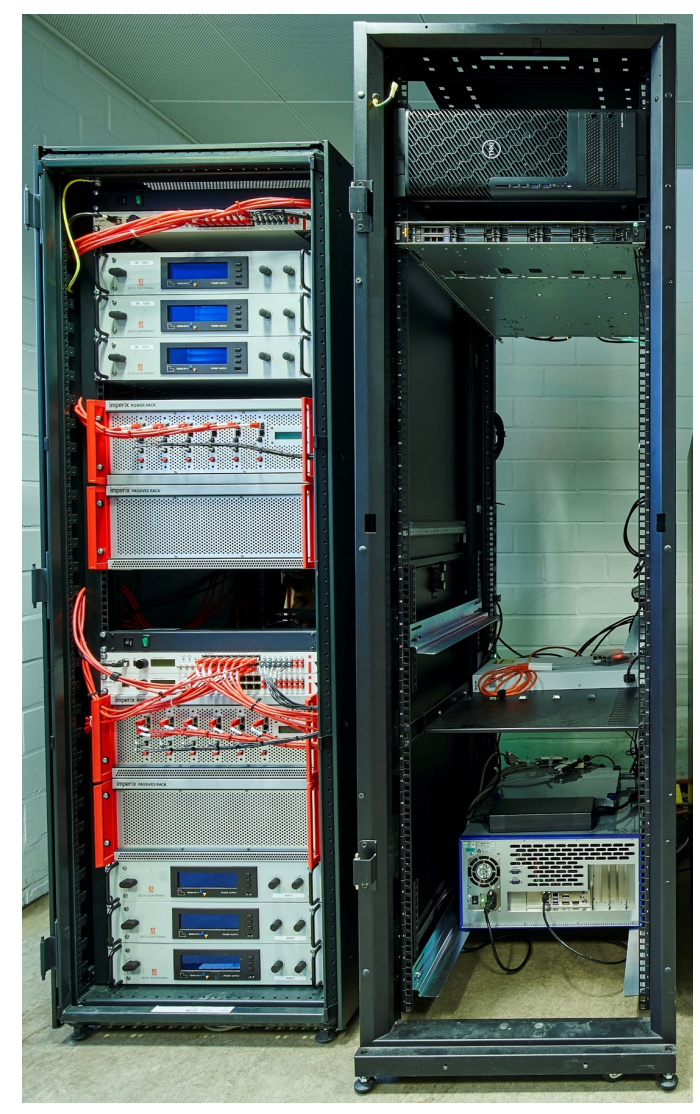

Fig. 3. Experimental setup with two GFMI units on the left, consisting of DC/DC and DC/AC conversion stages and controllers. The computation, communication, and measurement units are shown on the right.

and the transition variable is reset to $s(k) = 0$. During the transition, the variable is updated according to

$$s(k+1) = \min\left(s(k) + \frac{T_{\text{pc}}}{T_{\text{tr}}}, 1\right). \tag{11}$$

Hence, $s(k)$ increases from zero to one over the transition duration $T_{\text{tr}}$. The applied droop gain is then calculated as

$$K_{p1}(k) = K^0_{p1} + \lambda(s(k)) \left(K^{*0}_{p1} - K^0_{p1}\right). \tag{12}$$

Two standard interpolation profiles are considered for the transition. The cubic smoothstep profile

$$\lambda_3(s) = 3s^2 - 2s^3 \tag{13}$$

and the quintic smootherstep profile

$$\lambda_5(s) = 10s^3 - 15s^4 + 6s^5. \tag{14}$$

The cubic profile provides zero slope at the transition boundaries, while the quintic profile additionally provides zero curvature and therefore a smoother gain trajectory.

## IV. Case Study

The case study evaluates the dynamic behavior of the proposed transition mechanisms during adaptive active-power sharing. For this purpose, the active-power droop gain of $\text{GFMI}_2$ is changed during parallel operation, while its response is recorded. The investigated methods are compared with respect to the settling time of frequency as well as maximum frequency and power overshoot after the transient. The section is structured as follows. First, the experimental setup is described. Then, the investigated transition parameters and test procedure are introduced. Finally, the measured system responses are evaluated and compared.

TABLE I
CONTROL AND HARDWARE PARAMETERS

| Symbol | Value | Symbol | Value |
|---|---|---|---|
| $K_\mathrm{p}$, $K_\mathrm{q}$ | $2{\times}10^{-4}$, $1{\times}10^{-3}$ | $\omega^*$, $v_\mathrm{m}^*$ | $2\pi 50$ Hz, $230\sqrt{2}$ V |
| $C_\mathrm{g}$ | $10\,\mu$F | $L_\mathrm{g} = L_\mathrm{l}$ | 2.2 mH |
| $v_\mathrm{dc}^*$ | 700 V | $C_\mathrm{g}$, $f_\mathrm{s}$ | 20 kHz |
| $Z_\mathrm{L}$ | 3 kW | $P_1^*$, $P_2^*$ | 1 kW, 1 kW |

### A. Experimental Setup

The experimental setup extends the configuration presented in [15]. It consists of two identical 15 kW GFMI units, denoted as $\mathrm{GFMI}_1$ and $\mathrm{GFMI}_2$, connected in parallel as shown in Fig. 3. Each GFMI consists of three DC sources connected to a common DC bus through an interleaved boost converter, which regulates the DC bus voltage to 700 V. Further details on the DC bus control are provided in [16]. The controllers are programmed using ACG-SDK 2025.1.2 together with Simulink 2024b and are executed on imperix B-Box RCP 3.0 platforms [17], [18]. The converters operate at a switching frequency of $f_\mathrm{s} = 20$ kHz.

Higher-level coordination between the GFMI controllers and the distribution of reference values are enabled by a multimodal broker implemented through a software-defined overlay (SDO). The corresponding communication and MPC implementation are described in [8], [19]. Control measurements are acquired from the internal sensors of the parallel units, while evaluation data are recorded with a sampling rate of 200 kHz. The main control and hardware parameters are summarised in Table I. Here, $C_\mathrm{g}$ and $L_\mathrm{g}$ are the out capacitance and inductance of the inverter, respectively, and $L_\mathrm{l}$ is the line inductance of each GFMI.

### B. Test Procedure and Parameter Selection

The baseline configuration for the tests are two parallel GFMIs that supply a 3.15 kW load. As $P_1^* = P_2^* = 1$ kW and $K_\mathrm{p1} = K_\mathrm{p2}$, the remaining power offset is distributed evenly among both units. At time step $t_0 = 5$ s, the droop gain of $\mathrm{GFMI}_2$ is reduced by one order of magnitude to impose a new active power sharing ratio. The same initial condition and droop gain change is used for all investigated transition mechanisms to ensure comparability. The resulting responses are evaluated based on the maximum power overshoot $\Delta P_\mathrm{os}$ and maximum frequency overshoot $\Delta f_\mathrm{os}$ and frequency settling time $t_{\mathrm{set},f}$. The overshoot values are evaluated based on the final steady-state value, while the settling time is defined as the time required for the frequency to remain within a $10\,\%$ band around its final steady-state value. The event is repeated for different rate limits, low-pass filter cutoff frequencies and S-curve transition times. Furthermore, both cubic and quintic interpolation profiles are considered. The selected values cover conservative, nominal and aggressive settings for each transition mechanism. The evaluated parameters are summarised in Table II.

TABLE II
PARAMETER VALUES CONSIDERED FOR THE TRANSITION-MECHANISM CASE STUDY

| Method | Parameter | Values |
|---|---|---|
| Hard switching | – | – |
| Rate-limited transition | $m$ | $1{\times}10^{-4}$, $5{\times}10^{-4}$, $5{\times}10^{-3}$, $1{\times}10^{-2}$, $2{\times}10^{-2}$ |
| IIR low-pass filter | $f_c$ | 0.125 Hz, 0.25 Hz, 0.5 Hz, 1 Hz |
| S-curve transition | $\lambda(s)$ | cubic, quintic |
| S-curve transition | $T_\mathrm{tr}$ | 0.25 s, 0.5 s, 1 s, 2 s, 4 s |

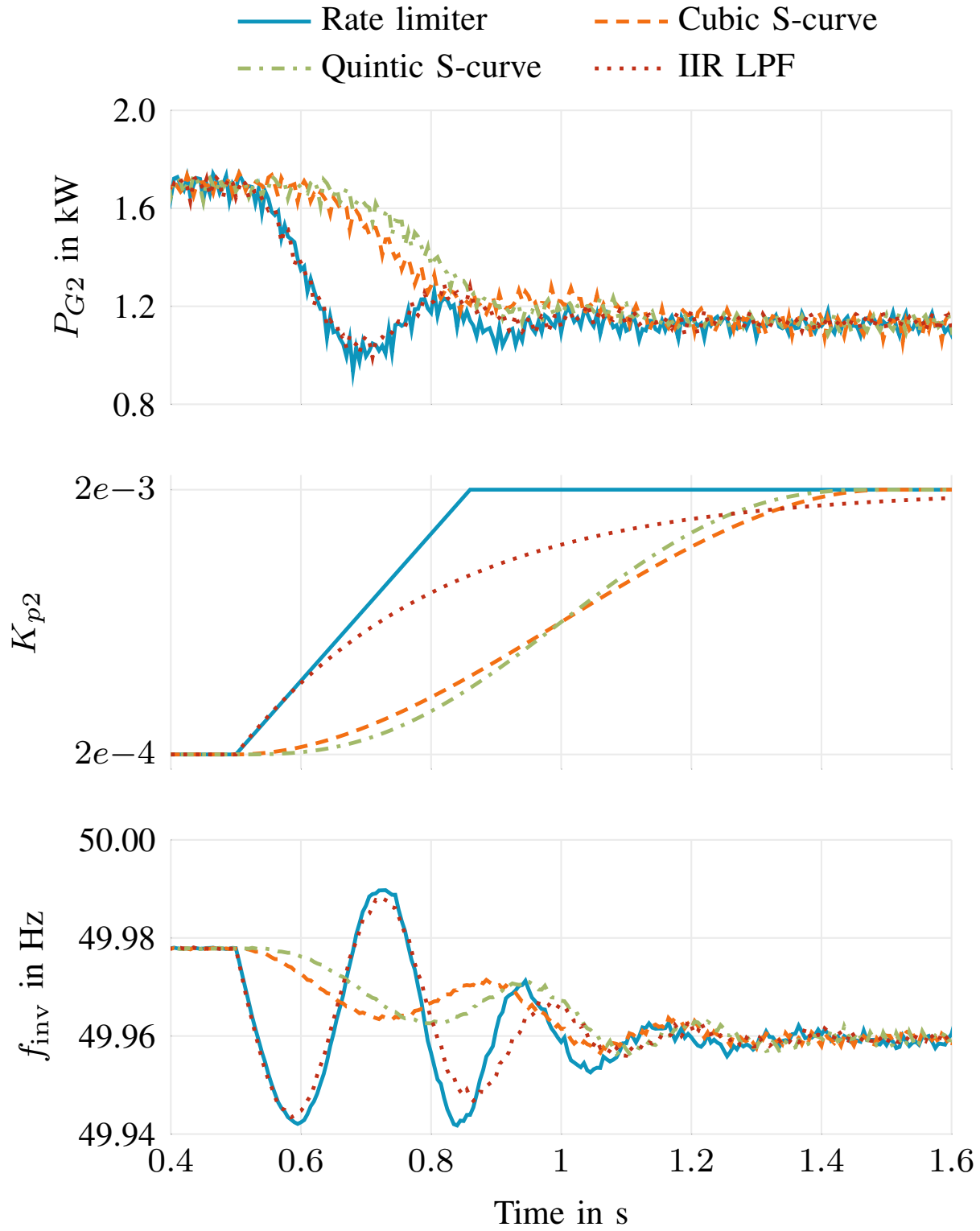

Fig. 4. Measured response of $\mathrm{GFMI}_2$ during the droop-gain transition for the best parameter setting of each investigated transition mechanism.

### C. Experimental results

Figure 4 shows the measured response of $\mathrm{GFMI}_2$ for the best parameter setting of each transition mechanism. The figure compares the active power $P_{G2}$, the applied droop gain $K_{p2}$ and the inverter frequency $f_\mathrm{inv}$ during the droop-gain transition. Table III summarises the best measured parameter setting for each transition mechanism. The overshoot values are evaluated with respect to the final steady-state value, while the settling time of the GFMI frequency is a $10\,\%$ band around the final steady-state value. For the Hard switching case, the largest transient response is produced, with an active power overshoot of 632.7 W and a frequency overshoot of 0.183 Hz.

TABLE III
BEST MEASURED PERFORMANCE OF THE INVESTIGATED TRANSITION MECHANISMS

| Method | Parameter | $\Delta P_{\mathrm{os}}$ | $t_{\mathrm{set},f}$ | $\Delta f_{\mathrm{os}}$ |
|---|---|---|---|---|
| Hard switching | – | $632.7\,\mathrm{W}$ | $0.681\,\mathrm{s}$ | $0.183\,\mathrm{Hz}$ |
| Rate limiter | $m = 5{\times}10^{-3}$ | $215.1\,\mathrm{W}$ | $0.779\,\mathrm{s}$ | $0.018\,\mathrm{Hz}$ |
| IIR LPF | $f_c = 0.5\,\mathrm{Hz}$ | $181.6\,\mathrm{W}$ | $0.703\,\mathrm{s}$ | $0.016\,\mathrm{Hz}$ |
| Cubic S-curve | $T_{\mathrm{tr}} = 1\,\mathrm{s}$ | $115.1\,\mathrm{W}$ | $0.709\,\mathrm{s}$ | $0.003\,\mathrm{Hz}$ |
| Quintic S-curve | $T_{\mathrm{tr}} = 1\,\mathrm{s}$ | $116.7\,\mathrm{W}$ | $0.747\,\mathrm{s}$ | $0.003\,\mathrm{Hz}$ |

This confirms that directly applying the updated droop gain can cause significant power and frequency transients. Both the rate-limited transition and the first-order IIR low-pass filter significantly reduce the transient response. The rate-limited transition with $m = 5{\times}10^{-3}$ reduces the active-power overshoot to $215.1\,\mathrm{W}$ and the frequency overshoot to $0.018\,\mathrm{Hz}$. The IIR low-pass filter with $f_c = 0.5\,\mathrm{Hz}$ provides a slightly lower active-power overshoot of $181.6\,\mathrm{W}$ and a comparable frequency overshoot of $0.016\,\mathrm{Hz}$.

The lowest transient deviations are achieved with the S-curve transitions. The cubic S-curve with $T_{\mathrm{tr}} = 1\,\mathrm{s}$ reduces the active-power overshoot to $115.1\,\mathrm{W}$ and the frequency overshoot to $0.003\,\mathrm{Hz}$. The quintic S-curve gives a similar result, with $116.7\,\mathrm{W}$ and $0.003\,\mathrm{Hz}$. Thus, both S-curve profiles provide the smoothest transition in the considered case study.

The settling times remain in a comparable range for all transition mechanisms. Hence, the main effect of the transition mechanisms is not a faster convergence to the final operating point, but a significant reduction of transient power and frequency overshoot. Overall, the S-curve transitions show the best transient mitigation, while the rate-limited transition and the IIR filter provide simpler alternatives with substantial improvement over hard switching.

## V. CONCLUSION

In this work transition mechanisms for adaptive active power sharing in parallel-operated GFMI were investigated. The objective was to apply updated droop-gain references from a supervisory layer without causing excessive power and frequency transients. Experimental results show that hard switching causes the largest transient response, with an active-power overshoot of $632.7\,\mathrm{W}$ and a frequency overshoot of $0.183\,\mathrm{Hz}$. All investigated transition mechanisms reduced these deviations. The strongest mitigation was achieved with the S-curve transitions, which reduced the active-power overshoot to approximately $115\,\mathrm{W}$ and limited the frequency overshoot to about $0.003\,\mathrm{Hz}$, while maintaining a similar settling time. Future work will integrate the transition mechanisms into a closed-loop MPC framework.